\documentclass[twocolumn,nofootinbib,superscriptaddress,floatfix]{revtex4-1}

\usepackage{array}

\usepackage[totalwidth=480pt, totalheight=680pt]{geometry}
\usepackage{setspace}

\usepackage{amsmath}
\usepackage{mathtools}
\usepackage{epstopdf}

\usepackage[english,greek]{babel}
\usepackage{verbatim}
\usepackage{booktabs}

\usepackage[usenames,dvipsnames]{color}
\usepackage{color}
\usepackage{pstricks,framed}

\usepackage{tensor}
\usepackage{amssymb}
\usepackage{amsthm}
\usepackage{amsfonts}
\usepackage{simplewick}

\usepackage{chngcntr}
\usepackage[retainorgcmds]{IEEEtrantools}
\usepackage{caption}

\usepackage{graphicx}
\usepackage[colorlinks=true, linktoc=page, citecolor=blue, urlcolor=blue]{hyperref}

\counterwithin{equation}{section}
\makeatletter
\def\p@subsection{}
\def\p@subsubsection{}
\makeatother

\allowdisplaybreaks

\begin{document}
\selectlanguage{english}

\title{\color{Blue}\textbf{M2-brane Dynamics in the Classical Limit of the BMN Matrix Model}}
\author{\textbf{Minos Axenides}}
\email{axenides@inp.demokritos.gr}
\affiliation{Institute of Nuclear and Particle Physics, N.C.S.R., "Demokritos",\\ 153 10, Agia Paraskevi, Greece}
\author{\textbf{Emmanuel Floratos}}
\email{mflorato@phys.uoa.gr}
\affiliation{Institute of Nuclear and Particle Physics, N.C.S.R., "Demokritos",\\ 153 10, Agia Paraskevi, Greece}
\affiliation{Department of Physics, National and Kapodistrian University of Athens,\\ Zografou Campus, 157 84, Athens, Greece}
\author{\textbf{Georgios Linardopoulos}}
\email{glinard@inp.demokritos.gr}
\affiliation{Institute of Nuclear and Particle Physics, N.C.S.R., "Demokritos",\\ 153 10, Agia Paraskevi, Greece}
\affiliation{Department of Physics, National and Kapodistrian University of Athens,\\ Zografou Campus, 157 84, Athens, Greece}

\begin{abstract}
\normalsize{\noindent We investigate the large-$N$ limit of the BMN matrix model by analyzing the dynamics of ellipsoidal M2-branes that spin in the 11-dimensional maximally supersymmetric $SO(3)\times SO(6)$ plane-wave background. We identify finite-energy solutions by specifying the local minima of the corresponding energy functional. These configurations are static in $SO(3)$ due to the Myers effect and rotate in $SO(6)$ with an angular momentum that is bounded from above. As a first step towards studying their chaotic properties, we evaluate the Lyapunov exponents of their radial fluctuations.}
\end{abstract}

\maketitle
\section[Introduction and Summary]{Introduction and Summary}
\noindent The study of chaotic phenomena in the vicinity of a black hole (BH) has attracted a lot of attention recently, mainly because of its close connection to the paradox of information loss \cite{ShenkerStanford13a, Hawking14, MaldacenaShenkerStanford15}. The observations of infalling observers (fifos) get scrambled by the microscopic degrees of freedom in the near-horizon region of the BH \cite{SusskindLindesay05, PapadodimasRaju12} and reach fiducial observers (fidos) in the form of chaotically processed information. Meanwhile the outgoing (soft+hard) Hawking radiation carries its own random correlation to the apparently lost information \cite{CarneyChauretteNeuenfeldSemenoff17, Strominger17}.\nocite{EllisMavromatosNanopoulos16b} \\[6pt]
\indent A very interesting proposal in the above framework consists in describing the chaotic and nonlocal dynamics of the BH horizon with a matrix model \cite{SekinoSusskind08}, specifically the matrix model of BFSS \cite{BFSS97} that can be considered as the Hamiltonian discretization of the BH membrane paradigm \cite{Damour78, ThornePriceMacdonald86}. A well-known property of the matrix model is that it reduces to a theory of supermembranes as the dimensionality $N$ of the corresponding matrices approaches infinity \cite{deWitHoppeNicolai88}. \\[6pt]
\indent In the present letter we initiate the systematic study of the chaotic properties of the large-$N$ limit of the BMN matrix model \cite{BMN02} (that is matrix theory on a plane-wave background) that is also equivalent to a theory of supermembranes \cite{DasguptaJabbariRaamsdonk02}. The stable fuzzy sphere solutions of the BMN matrix model hopefully describe the BH horizon geometry and can be used for the study of its fluctuations \cite{Gur-AriHanadaShenker15, AsanoKawaiYoshida15}. Here we focus on a specific ansatz that consists of a spinning ellipsoid in the 11-dimensional maximally supersymmetric plane-wave background. Our system is introduced in full generality in \S\ref{Section:GeneralSetup}. In \S\ref{Section:ParticularSolutions} we discuss some of the simplest possible solutions and in \S\ref{Section:StabilityAnalysis} we examine their radial stability.
\section[General Setup]{General Setup \label{Section:GeneralSetup}}
\noindent The Hamiltonian of a bosonic relativistic membrane in the 11-dimensional maximally supersymmetric plane-wave background,
\begin{IEEEeqnarray}{l}
ds^2 = -2 dx^{+} dx^{-} + \sum_{i=1}^3 dx_i dx_i + \sum_{j=1}^6 dy_j dy_j - \nonumber \\
\hspace{1cm}- \left[\frac{\mu^2}{9}\sum_{i=1}^3 x_i x_i + \frac{\mu^2}{36}\sum_{j=1}^6 y_j y_j\right] dx^+ dx^+ \qquad \label{MaximallySupersymmetricMetric} \\
F_{123+} = \mu \label{MaximallySupersymmetricMetricFieldStrength}
\end{IEEEeqnarray}
reads, in the so-called light-cone gauge $x^+ = \tau$ \cite{DasguptaJabbariRaamsdonk02}:
\begin{IEEEeqnarray}{ll}
H = \frac{T}{2}\int_{\Sigma} d^2\sigma\bigg[p_x^2 + p_y^2 + \frac{1}{2}\left\{x_i,x_j\right\}^2 + \frac{1}{2}\left\{y_i,y_j\right\}^2 + \nonumber \\
+ \left\{x_i,y_j\right\}^2 + \frac{\mu^2 x^2}{9} + \frac{\mu^2 y^2}{36} - \frac{\mu}{3}\,\epsilon_{ijk}\left\{x_i,x_j\right\}x_k\bigg], \qquad \label{ppWaveHamiltonian}
\end{IEEEeqnarray}
where the indices of the coordinates $x$ run from 1 to 3 while those of $y$ run from 1 to 6. In this gauge \eqref{ppWaveHamiltonian} has a residual invariance under (time-independent) area-preserving diffeomorphisms SDiff$\left(\Sigma\right)$, generated by the Gauss law constraint:
\begin{IEEEeqnarray}{l}
\left\{\dot{x}_i, x_i\right\} + \left\{\dot{y}_j, y_j\right\} = 0. \label{GaussLaw}
\end{IEEEeqnarray}
The equations of motion for the spatial coordinates $x$ and $y$ that are derived from the Hamiltonian \eqref{ppWaveHamiltonian} are given by:
\begin{IEEEeqnarray}{ll}
\ddot{x}_i = &\left\{\left\{x_i,x_j\right\},x_j\right\} + \left\{\left\{x_i,y_j\right\},y_j\right\} - \frac{\mu^2}{9}\,x_i + \nonumber \\
& +\frac{\mu}{2}\epsilon_{ijk}\left\{x_j,x_k\right\} \label{xEquation} \\
\ddot{y}_i = &\left\{\left\{y_i,y_j\right\},y_j\right\} + \left\{\left\{y_i,x_j\right\},x_j\right\} - \frac{\mu^2}{36}\,y_i. \qquad \label{yEquation}
\end{IEEEeqnarray}
\indent In the case of spherical membrane topologies that will be discussed in this letter, the appropriate set of functions describing their internal degrees of freedom are the well-known spherical harmonics $Y_{jm}\left(\theta,\phi\right)$ ($j = 0,1,\ldots$, $\left|m\right| = 0,1,\ldots j$). $Y_{jm}\left(\theta,\phi\right)$ satisfy the infinite-dimensional Lie algebra SDiff$\left(\text{S}^2\right)$ \cite{Hoppe82}:
\begin{IEEEeqnarray}{c}
\left\{Y_{j_1 m_1},Y_{j_2 m_2}\right\} = f_{j_1 m_1, j_2 m_2}^{j_3 m_3} Y_{j_3 m_3},
\end{IEEEeqnarray}
and are harmonic and homogeneous polynomials of the coordinates $e_i$:
\begin{IEEEeqnarray}{c}
(e_1, e_2, e_3) = (\cos\phi \sin\theta, \sin\phi \sin\theta, \cos\theta) \nonumber \\
\phi \in [0,2\pi), \quad \theta \in [0,\pi], \label{Epsilon1}
\end{IEEEeqnarray}
which satisfy the $\mathfrak{so}\left(3\right)$ Poisson algebra,
\begin{IEEEeqnarray}{c}
\{e_a, e_b\} = \epsilon_{abc} \, e_c, \quad \int e_a \, e_b\,d^2\sigma = \frac{4\pi}{3} \, \delta_{ab} \qquad \ \label{Epsilon2}
\end{IEEEeqnarray}
and are orthonormal. The spatial coordinates $x$ and $y$ can be expanded in spherical harmonics as
\begin{IEEEeqnarray}{c}
x_i = \sum_{j,m} x_{i}^{jm}\left(\tau\right) Y_{jm}\left(\theta,\phi\right) \\
y_i = \sum_{j,m} y_{i}^{jm}\left(\tau\right) Y_{jm}\left(\theta,\phi\right),
\end{IEEEeqnarray}
which leads to an infinite system of coupled second order ODEs for the mode functions $x_{i}^{jm}\left(\tau\right)$ and $y_{i}^{jm}\left(\tau\right)$. For consistency, the initial values of the mode functions and their time derivatives should satisfy the Gauss-law constraint \eqref{GaussLaw}. \\[6pt]
\indent Now it is known that the only finite subalgebra of SDiff$\left(\text{S}^2\right)$ that can be used to reduce the aforementioned infinite system of equations to a finite system is $\mathfrak{so}\left(3\right)$ \cite{Banyaga78}. In light of this, let us consider the following $\mathfrak{so}\left(3\right)$-invariant ansatz that automatically satisfies the Gauss-law constraint \eqref{GaussLaw}:
\begin{IEEEeqnarray}{ll}
x_i = \tilde{u}_i\left(\tau\right) e_i, \quad & y_j = \tilde{v}_j\left(\tau\right) e_j, \label{Ansatz1} \\[6pt]
& y_{j+3} = \tilde{w}_j\left(\tau\right) e_j, \ i,j = 1,2,3. \qquad \quad \label{Ansatz2}
\end{IEEEeqnarray}
The reduced system for $(\tilde{u}_i, \tilde{v}_i, \tilde{w}_i)$ is an interesting dynamical system with stable and unstable solutions corresponding to rotating and pulsating membranes of spherical topology. The ansatz \eqref{Ansatz1}--\eqref{Ansatz2} leads to the Hamiltonian:
\begin{IEEEeqnarray}{l}
H = \frac{2\pi T}{3}\left(\tilde{p}_u^2 + \tilde{p}_v^2 + \tilde{p}_w^2\right) + U, \label{Hamiltonian}
\end{IEEEeqnarray}
obtained by integrating the internal coordinates $\theta$ and $\phi$. The potential energy $U$ is given by
\begin{IEEEeqnarray}{ll}
U = &\frac{2\pi T}{3}\bigg[\tilde{u}_1^2 \tilde{u}_2^2 + \tilde{u}_2^2 \tilde{u}_3^2 + \tilde{u}_3^2 \tilde{u}_1^2 + \tilde{r}_1^2 \tilde{r}_2^2 + \tilde{r}_2^2 \tilde{r}_3^2 + \tilde{r}_3^2 \tilde{r}_1^2 + \nonumber \\
& + \tilde{u}_1^2 \left(\tilde{r}_2^2 + \tilde{r}_3^2\right) + \tilde{u}_2^2 \left(\tilde{r}_3^2 + \tilde{r}_1^2\right) + \tilde{u}_3^2 \left(\tilde{r}_1^2 + \tilde{r}_2^2\right) + \nonumber \\
& + \frac{\mu^2}{9}\left(\tilde{u}_1^2 + \tilde{u}_2^2 + \tilde{u}_3^2\right) + \frac{\mu^2}{36}\left(\tilde{r}_1^2 + \tilde{r}_2^2 + \tilde{r}_3^2\right) - \nonumber \\
& - 2\mu \tilde{u}_1 \tilde{u}_2 \tilde{u}_3 \bigg], \quad \tilde{r}_j^2 \equiv \tilde{v}_j^2 + \tilde{w}_j^2, \ j = 1,2,3. \label{Potential1}
\end{IEEEeqnarray}
\indent The Hamiltonian \eqref{Hamiltonian} has an obvious $SO(2)\times SO(2)\times SO(2)$ symmetry in the coordinates $\tilde{v}_i$ and $\tilde{w}_i$ so that any solution will preserve three $SO(2)$ angular momenta $\ell_i$ ($i = 1,2,3$). In terms of the conserved momenta $\ell_i$ the kinetic terms of \eqref{Hamiltonian} can be written as follows:
\begin{IEEEeqnarray}{l}
\tilde{p}_v^2 + \tilde{p}_w^2 = \sum_{i=1}^3\left(\dot{\tilde{r}}_i^2 + \frac{\ell_i^2}{\tilde{r}_i^2}\right)
\end{IEEEeqnarray}
leading to the effective potential
\begin{IEEEeqnarray}{ll}
V_{\text{eff}} = U + \frac{2\pi T}{3}\left(\frac{\ell_1^2}{\tilde{r}_1^2} + \frac{\ell_2^2}{\tilde{r}_2^2} + \frac{\ell_3^2}{\tilde{r}_3^2}\right). \label{EffectivePotential}
\end{IEEEeqnarray}
\indent The effective potential \eqref{EffectivePotential} has four distinct types of terms: $\bullet$ (1) angular momentum terms (repulsive), $\bullet$ (2) quartic interaction terms (attractive), $\bullet$ (3) mass terms (attractive) and $\bullet$ (4) cubic Myers terms (repulsive). The last two types of terms are $\mu$-dependent and are thus absent in the flat space limit ($\mu \rightarrow 0$) that was studied in \cite{AxenidesFloratos07}. The presence of two extra repulsive and attractive terms for $\mu \neq 0$ (due to the plane-wave background) increases the richness of the resulting system, as it will become apparent below.
\section[Simplest Solutions]{Simplest Solutions \label{Section:ParticularSolutions}}
\noindent There are many known solutions of the BMN matrix model and its classical ($N\rightarrow \infty$) limit that is the membrane in the plane-wave background \eqref{MaximallySupersymmetricMetric}--\eqref{MaximallySupersymmetricMetricFieldStrength}. BPS solutions of various topologies have been studied in \cite{Bak02a, Mikhailov02b, Park02b, BakKimLee05, HoppeLee07}, while many rotating (non-BPS) solutions have been found in \cite{ArnlindHoppe03b, ArnlindHoppeTheisen04, BerensteinDzienkowskiLashof-Regas15, Hoppe15}. Below we identify bouncing membrane solutions and (from the critical points of the effective potential \eqref{EffectivePotential}) rotating solutions.
\subsection[$SO\left(3\right)$ Sector]{$SO\left(3\right)$ Sector}
\noindent Let us first consider the $SO\left(3\right)$ sector that is obtained by setting the $SO\left(6\right)$ variables $\tilde{v}_i$ and $\tilde{w}_i$ equal to zero. If we scale out $\mu$ (i.e.\ set $x_i \equiv \mu u_i e_i$) the effective potential of the membrane becomes:
\begin{IEEEeqnarray}{ll}
V_{\text{eff}} = &\frac{2\pi T \mu^4}{3}\bigg[u_1^2 u_2^2 + u_2^2 u_3^2 + u_1^2 u_3^2 + \nonumber \\
& + \frac{1}{9}\left(u_1^2 + u_2^2 + u_3^2\right) - 2 u_1 u_2 u_3\bigg], \label{StaticPotential}
\end{IEEEeqnarray}
that is also known as the generalized 3-dimensional H\'{e}non-Heiles potential. We can determine the simplest critical points of \eqref{StaticPotential} and then all the others can be obtained by flipping the sign of exactly two out of three $u_i$'s. We get $\textbf{u}_0 = 0$ and
\begin{IEEEeqnarray}{ll}
\textbf{u}_{1/6} = \frac{1}{6}\cdot\left(1, 1, 1\right), \quad \textbf{u}_{1/3} = \frac{1}{3}\cdot\left( 1, 1, 1\right). \qquad \label{So3Extrema1}
\end{IEEEeqnarray}
The effective potential \eqref{StaticPotential} has the symmetry of a tetrahedron $T_d$ formed by the four critical points \eqref{So3Extrema1}. There are two degenerate minima at $\textbf{u}_0$ (a point-like membrane) and $\textbf{u}_{1/3}$ (the Myers dielectric sphere), and a saddle point at $\textbf{u}_{1/6}$:
\begin{IEEEeqnarray}{c}
V_{\text{eff}}\left(0\right) = V_{\text{eff}}\left(\frac{1}{3}\right) = 0,\quad V_{\text{eff}}\left(\frac{1}{6}\right) = \frac{2\pi T \mu^{4}}{6^4}. \qquad \label{So3Extrema2}
\end{IEEEeqnarray}
\indent When the $u_i$ are not all equal, the equations of motion have a complicated form so that the time-dependent solutions can only be found numerically. For $u_1 = u_2 = u_3$ the problem reduces to the exactly solvable case of the double-well potential (cf.\ \cite{BrizardWestland17}). Let us briefly present the explicit solutions that are periodically bouncing spherical membranes in just one or both lobes of the double-well potential. \\
%
%\begin{comment}
\begin{center}
\includegraphics[scale=0.4]{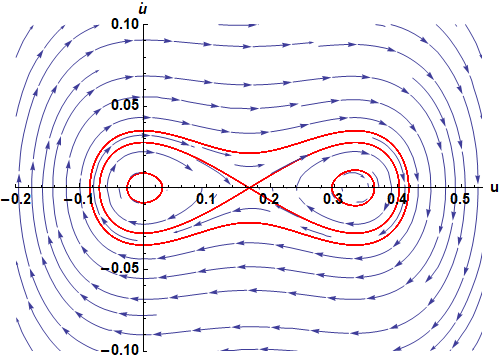}
\captionof{figure}{Phase portrait of the $SO(3)$ membrane.} \label{Graph:StreamPlot}
\end{center}
%\end{comment}

\indent For $u = u_1 = u_2 = u_3$ the Hamiltonian of the membrane becomes:
\begin{IEEEeqnarray}{c}
H = 2\pi T\mu^4 \left[p^2 + u^2\left(u - \frac{1}{3}\right)^2\right] \qquad \label{So3Hamiltonian}
\end{IEEEeqnarray}
implying the following equations of motion:
\begin{IEEEeqnarray}{c}
\dot{u} = p, \qquad \dot{p} = -u\left(2u^2 - u + \frac{1}{9}\right), \qquad \label{So3Equations}
\end{IEEEeqnarray}
where we switch to dimensionless time $t \equiv \mu \, \tau$ from now on. The phase portrait of the system \eqref{So3Equations} has been drawn in figure \ref{Graph:StreamPlot}. There are three kinds of orbits: $\bullet$ (1) oscillations of small energies ($\mathcal{E} \equiv E/ 2\pi T\mu^4 < 6^{-4} \equiv \mathcal{E}_{c}$) around either of the two stable global minima ($u_0 = 0, 1/3$), $\bullet$ (2) oscillations of larger energies ($\mathcal{E} > \mathcal{E}_{c}$) around the local maximum ($u_0 = 1/6$) and $\bullet$ (3) two homoclinic orbits through the unstable equilibrium point at $u_0 = 1/6$ with energy equal to the potential height ($\mathcal{E} = \mathcal{E}_{c}$). \\[6pt]
\indent The expressions for the orbits can be computed from the energy integral and the initial conditions
\begin{IEEEeqnarray}{l}
\dot{u}_0\left(0\right) = 0, \qquad u_0\left(0\right) = \frac{1}{6} \pm \sqrt{\frac{1}{6^2} + \sqrt{\mathcal{E}}}, \qquad
\end{IEEEeqnarray}
where the plus/minus signs correspond to the right/left side of the double-well potential. We find:
\begin{IEEEeqnarray}{ll}
u_0\left(t\right) = &\frac{1}{6} \pm \sqrt{\frac{1}{6^2} + \sqrt{\mathcal{E}}} \cdot \nonumber \\
&\cdot cn\left[\sqrt{2\sqrt{\mathcal{E}}}\cdot t\Bigg|\frac{1}{2}\left(1 + \frac{1}{36\sqrt{\mathcal{E}}}\right)\right]. \qquad \ \label{OrbitSO3}
\end{IEEEeqnarray}
For $\mathcal{E} \geq \mathcal{E}_{c}$ only the plus sign should be kept in \eqref{OrbitSO3}. For the critical energy $\mathcal{E} = \mathcal{E}_{c}$, \eqref{OrbitSO3} reduces to the homoclinic orbit:
\begin{IEEEeqnarray}{c}
u_0\left(t\right) = \frac{1}{6} \pm \frac{1}{3 \sqrt{2}} \cdot \text{sech}\left(\frac{t}{3 \sqrt{2}}\right). \qquad \label{HomoclinicOrbitSO3}
\end{IEEEeqnarray}
\indent The plot of \eqref{OrbitSO3}--\eqref{HomoclinicOrbitSO3} for various values of the energy $\mathcal{E}$ can be found in figures \ref{Graph:OrbitsSO3a}--\ref{Graph:OrbitsSO3c}. The lower plot of figure \ref{Graph:OrbitsSO3a} describes single-well oscillations of the membrane around the point-like configuration, whereas the upper plot describes oscillations around the Myers sphere. Because of the potential barrier, the latter cannot shrink the membrane to a point as it happens in the former case or for $\mathcal{E} > \mathcal{E}_c$. Note also that for $u < 0$ the orientation of the membrane is reversed.
%
%\begin{comment}
\begin{center}
\includegraphics[scale=0.31]{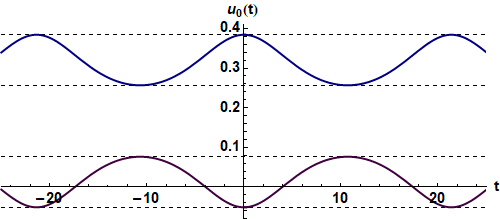}
\captionof{figure}{Plot of \eqref{OrbitSO3} for $\mathcal{E} < 1/6^4$.} \label{Graph:OrbitsSO3a}
\includegraphics[scale=0.31]{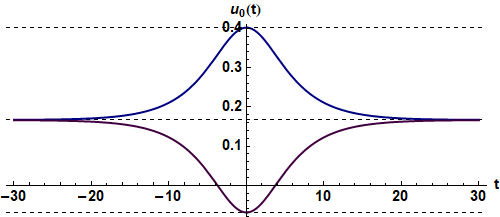}
\captionof{figure}{Plot of \eqref{HomoclinicOrbitSO3} or \eqref{OrbitSO3} for $\mathcal{E} = 1/6^4$.} \label{Graph:OrbitsSO3b}
\includegraphics[scale=0.31]{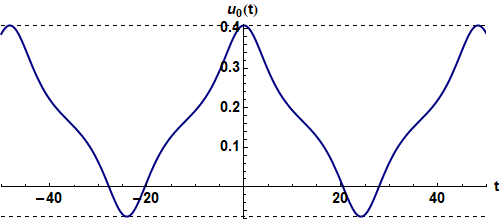}
\captionof{figure}{Plot of \eqref{OrbitSO3} for $\mathcal{E} > 1/6^4$.} \label{Graph:OrbitsSO3c}
\end{center}
%\end{comment}

\indent The period as a function of the energy is given in terms of the complete elliptic integral of the first kind:
\begin{IEEEeqnarray}{ll}
T\left(\mathcal{E}\right) = 2\sqrt{\frac{2}{\sqrt{\mathcal{E}}}} \cdot \textbf{K}\left(\frac{1}{2}\left(1 + \frac{1}{36\sqrt{\mathcal{E}}}\right)\right) \label{PeriodSO3}
\end{IEEEeqnarray}
and it has been plotted in figure \ref{Graph:PeriodSO3}. The period of the homoclinic orbit \eqref{HomoclinicOrbitSO3} is infinite.
\begin{center}
\includegraphics[scale=0.31]{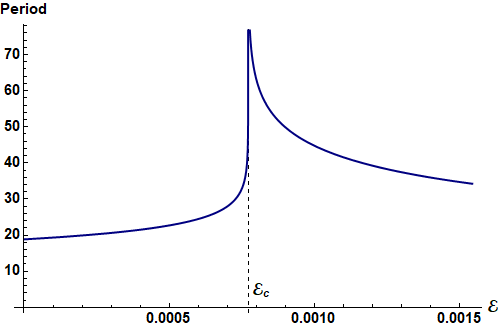}
\captionof{figure}{Period \eqref{PeriodSO3} as a function of energy.} \label{Graph:PeriodSO3}
\end{center}
\subsection[$SO\left(3\right)\times SO\left(6\right)$ Sector]{$SO\left(3\right)\times SO\left(6\right)$ Sector}
\noindent Let us now consider the simplest axially symmetric configuration that extends in the full geometric background of $SO(3)\times SO(6)$. This configuration consists of a membrane that is static in the $SO(3)$ sector and rigidly rotating in $SO(6)$:
\begin{IEEEeqnarray}{ll}
u_i \equiv \mu u\left(t\right), \quad &v_j \equiv \mu v\left(t\right)\cos\varphi\left(t\right) \label{Ansatz3} \\
& w_j \equiv \mu v\left(t\right)\sin\varphi\left(t\right), \qquad \label{Ansatz4}
\end{IEEEeqnarray}
where $i,j= 1,2,3$.\footnote{Note the similarity between \eqref{Ansatz3}--\eqref{Ansatz4} and the definition of the cylindrical coordinate system, for $(z,\rho) = \mu(u, v)$.} With the ansatz \eqref{Ansatz3}--\eqref{Ansatz4} the Hamiltonian \eqref{Hamiltonian}--\eqref{EffectivePotential} becomes:
\begin{IEEEeqnarray}{ll}
\frac{H}{2 \pi T \mu^4} = p_u^2 + p_v^2 + V
\end{IEEEeqnarray}
where
\begin{IEEEeqnarray}{ll}
V \equiv \frac{V_{\text{eff}}}{2 \pi T \mu^4} = u^4 &+ 2 u^2 v^2 + v^4 + \frac{u^2}{9} + \frac{v^2}{36} - \qquad \nonumber \\
&- \frac{2 u^3}{3} + \frac{\ell ^2}{v^2} \qquad \label{AxiallySymmetricPotential}
\end{IEEEeqnarray}
and the conserved angular momentum is scaled as
\begin{IEEEeqnarray}{l}
\ell \mu^3 \equiv \ell_{1} = \ell_{2} = \ell_{3}. \qquad \label{Ansatz5}
\end{IEEEeqnarray}
The equations of motion read ($p_u = \dot{u}$, $p_v = \dot{v}$):
\begin{IEEEeqnarray}{ll}
\ddot{u} = -u \left[2 u^2 - u + \frac{1}{9} + 2 v^2\right] \qquad \label{AxiallySymmetricExtrema1} \\
\ddot{v} = -\frac{1}{v^3} \left[2v^6 + \left(\frac{1}{36} + 2u^2\right)v^4 -\ell^2\right]. \qquad \label{AxiallySymmetricExtrema2}
\end{IEEEeqnarray}
\indent We now proceed to the study of the critical points of \eqref{AxiallySymmetricPotential} that are found by solving \eqref{AxiallySymmetricExtrema1}--\eqref{AxiallySymmetricExtrema2} at the equilibrium points $\ddot{u} = \ddot{v} = 0$, where $u = u_0$ and $v = v_0$ are constant. To satisfy \eqref{AxiallySymmetricExtrema1}, we should either have $u_0 = 0$ or the quantity
\begin{IEEEeqnarray}{ll}
v_0^2 = \left(u_0 - \frac{1}{6}\right)\left(\frac{1}{3} - u_0\right) > 0 \label{AxiallySymmetricExtrema3}
\end{IEEEeqnarray}
must be positive.
For $u \neq 0$ \eqref{AxiallySymmetricExtrema3} leads to the following bounds on the allowed values of $u_0$ and $v_0$:
\begin{IEEEeqnarray}{ll}
\frac{1}{6} \leq u_0 \leq \frac{1}{3} \quad \& \quad 0 \leq v_0 \leq \frac{1}{12} \equiv v_{\text{max}}. \qquad \label{ExtremalBounds}
\end{IEEEeqnarray}
\indent The second equilibrium condition \eqref{AxiallySymmetricExtrema2} implies for $\ell \neq 0$, $\dot{\varphi} = \omega$ (constant) and $\ddot{v} = 0$:
\begin{IEEEeqnarray}{l}
\omega^2 = 2u_0^2 + 2v_0^2 + \frac{1}{36} \qquad \& \qquad \ell = \omega v_0^2. \qquad \label{AngularVelocity}
\end{IEEEeqnarray}
Inserting \eqref{AxiallySymmetricExtrema3} into \eqref{AngularVelocity} we can express the conserved angular momentum in terms of $u_0$:
\begin{IEEEeqnarray}{ll}
\ell^2 = \left(u_0 - \frac{1}{12}\right)\left(u_0 - \frac{1}{6}\right)^2\left(\frac{1}{3} - u_0\right)^2 \qquad \label{BinaryAngularMomentum}
\end{IEEEeqnarray}
and similarly for the energy \eqref{AxiallySymmetricPotential}:
\begin{IEEEeqnarray}{ll}
\mathcal{E} = \frac{5}{3}\left(u_0 - \frac{1}{12}\right)\left(u_0 - \frac{2}{15}\right)\left(\frac{1}{3} - u_0\right), \qquad \label{BinaryEnergy}
\end{IEEEeqnarray}
which is positive inside the range \eqref{ExtremalBounds}. The plot of \eqref{BinaryEnergy} has been drawn with a red dashed line in figure \ref{Graph:EffectivePotential} where we have also plotted \eqref{AxiallySymmetricPotential} for various $v_0$'s.
\begin{center}
\includegraphics[scale=0.31]{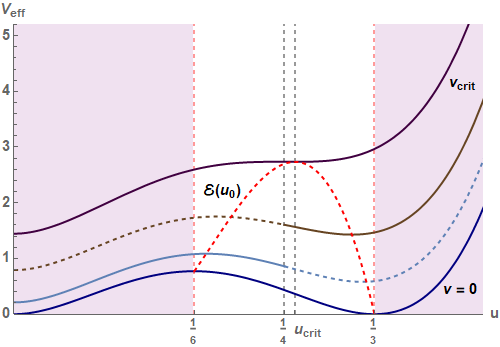}
\captionof{figure}{\eqref{AxiallySymmetricPotential} for various $v$'s and $\ell$'s.} \label{Graph:EffectivePotential}
\end{center}

\indent It is obvious from the expressions \eqref{BinaryAngularMomentum}--\eqref{BinaryEnergy} that both the energy and the angular momentum have a maximum that occurs at the same value of $u_0 \neq 0$ inside the physical region \eqref{ExtremalBounds}:
\begin{IEEEeqnarray}{l}
u_{\text{crit}} = \frac{1}{60} \left(11+\sqrt{21}\right) \approx 0.25971 \\[6pt]
v_{\text{crit}} = \frac{1}{30} \sqrt{2\sqrt{21} - 3} \approx 0.0827657. \label{AxiallySymmetricExtrema4}
\end{IEEEeqnarray}
\begin{center}
\includegraphics[scale=0.31]{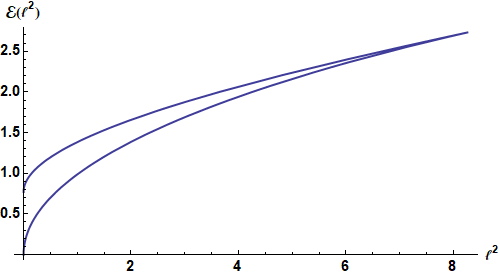}
\captionof{figure}{Dispersion relation $\mathcal{E} = \mathcal{E}\left(\ell^2\right)$.} \label{Graph:DispersionRelation}
\end{center}
This explains the cusp in the dispersion relation $\mathcal{E} = \mathcal{E}\left(\ell^2\right)$ (see figure \ref{Graph:DispersionRelation}). For $u = 0$ the system reduces to an Euler-top membrane in $SO\left(6\right)$. These configurations have been studied in \cite{AxenidesFloratos07} and have no bound in either the energy or the angular momentum. \\[6pt]
\indent In order to specify the type of each critical point of \eqref{AxiallySymmetricExtrema1}--\eqref{AxiallySymmetricExtrema2}, let us evaluate the $2\times 2$ Hessian matrix:
\begin{IEEEeqnarray}{l}
H = \left(\begin{array}{cc} 2u_0\left(4u_0-1\right) & 8u_0v_0 \\ 8u_0v_0 & -8u_0^2 + 12u_0 - 10/9\end{array}\right). \qquad \ \label{HessianMatrix}
\end{IEEEeqnarray}
From the eigenvalues of the Hessian \eqref{HessianMatrix} we find two sets of critical points: a series of saddle points between $1/6 \leq u_0 \leq u_{\text{crit}}$ and a series of minima between $u_{\text{crit}} < u_0 \leq 1/3$. Inverting \eqref{AxiallySymmetricExtrema3} we get
\begin{IEEEeqnarray}{ll}
u_{\pm} = \frac{1}{4} \pm \sqrt{v_{\text{max}}^2 - v_0^2}, \label{MaximaMinima}
\end{IEEEeqnarray}
where $u_-$ parametrizes the series of saddle points between $1/6 \leq u_0 \leq 1/4$ and $u_+$ parameterizes the series of minima between $u_{\text{crit}} < u_0 \leq 1/3$ and the series of saddle points between $1/4 \leq u_0 \leq u_{\text{crit}}$. The former reduces to the unstable point $\textbf{u}_{1/6}$ of the double-well potential when the $SO(6)$ coordinate $v$ becomes zero while the latter reduce to the Myers minimum $\textbf{u}_{1/3}$. For $v_0 > 0$ the degeneracy \eqref{So3Extrema2} of the double-well at $u_0 =0, 1/3$ is broken and the two critical points at $\textbf{u}_{1/6}$ and $\textbf{u}_{1/3}$ rise towards $u_0 = 1/4$ with
\begin{IEEEeqnarray}{ll}
\mathcal{E}_+ - \mathcal{E}_- = \frac{10}{3}\left(v_0^2 - \frac{1}{360}\right)\sqrt{v_{\text{max}}^2 - v_0^2}, \qquad \label{PotentialDifference}
\end{IEEEeqnarray}
where $\mathcal{E}_{\pm} \equiv \mathcal{E}\left(u_{\pm}\right)$. Notice that the minima $u_+$ are energetically favored only inside the interval $0 \leq v_0 \leq 1/6\sqrt{10} < v_{\text{crit}}$, while for $1/6\sqrt{10} < v_0 \leq v_{\text{crit}}$ the minima $u_+$ have larger energies than the saddle points $u_-$. At $v = v_{\text{max}}$, $u_+ = u_- = 1/4$ the difference \eqref{PotentialDifference} vanishes and the two series of saddle points $u_{\pm}$ coalesce. Beyond the critical values of $u$ and $v$ there is no balancing of the forces and the motion of the membrane can become chaotic.\footnote{See e.g.\ \cite{AsanoKawaiYoshida15} for a study of the dynamical system that emerges in the case $\ell = 0$.}
\section[Stability Analysis]{Stability Analysis \label{Section:StabilityAnalysis}}
\noindent In this section we will examine the radial stability of the above membrane configurations. The angular stability can be studied along the lines of \cite{AxenidesFloratosPerivolaropoulos00, AxenidesFloratosPerivolaropoulos01} and will be the subject of a forthcoming work \cite{AxenidesFloratosLinardopoulos17b}. \\[6pt]
\indent Let us begin with the static membrane in $SO(3)$ that we discussed at the beginning of the previous section. The nine critical points of the $SO(3)$ potential \eqref{StaticPotential} have been given in \eqref{So3Extrema1}. It is easy to show that the corresponding Hessian matrix is positive-definite for $\textbf{u}_0$ and $\textbf{u}_{1/3}$ and indefinite for $\textbf{u}_{1/6}$. Therefore the former are (global) minima of the potential and the latter is a saddle point. \\[6pt]
\indent The same conclusion can be drawn by perturbing the corresponding equations of motion and transforming the resulting linearized system into an eigenvalue/eigenvector problem. We find the following eigenvalues for each of the nine critical points:
%
%\begin{comment}
\vspace{-1cm}\begin{center}
\begin{eqnarray}
\begin{array}{|c|c|c|}
\hline && \\
\text{extremum}& \text{eigenvalues } \lambda^2 \text{ (\#)} & \text{stability} \\[6pt]
\hline && \\
\textbf{u}_0 & -\frac{1}{9} \, \left(3\right), \ -\frac{1}{36} \, \left(6\right) & \text{center (S)} \\[12pt]
\textbf{u}_{1/6} & \frac{1}{18} \, \left(1\right), \ -\frac{5}{18} \, \left(2\right), \ -\frac{1}{12} \, \left(6\right) & \text{saddle point} \\[12pt]
\textbf{u}_{1/3} & -\frac{1}{9} \, \left(1\right), \ -\frac{4}{9} \, \left(2\right), \ -\frac{1}{4} \, \left(6\right) & \text{center (S)} \\[6pt]
\hline
\end{array} \nonumber
\end{eqnarray}
\end{center}
%\end{comment}

\begin{figure}
\begin{center}
\includegraphics[scale=0.4]{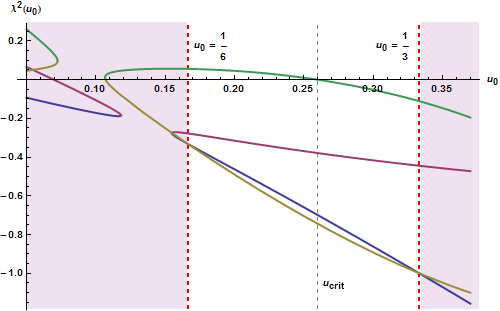}
\caption{Plot of the eigenvalues \eqref{AxiallySymmetricEigenvalues1}--\eqref{AxiallySymmetricEigenvalues2} as a function of the coordinate $u_0$.} \label{Graph:Eigenvalues}
\end{center}
\end{figure}

\indent Each negative eigenvalue corresponds to a stable direction, whereas the positive eigenvalues give rise to stable/unstable directions, depending on the sign of the real eigenvalue $\lambda$. This confirms the existence of two stable degenerate (global) minima ($\textbf{u}_0$ and $\textbf{u}_{1/3}$) and a single saddle point ($\textbf{u}_{1/6}$) between them. \\[6pt]
\indent Let us now treat the case of the $SO(3)\times SO(6)$ dielectric membrane \eqref{Ansatz3}--\eqref{Ansatz4}. Here's the solution of the equations of motion \eqref{AxiallySymmetricExtrema1}--\eqref{AxiallySymmetricExtrema2} (for $i,j = 1,2,3$):
\begin{IEEEeqnarray}{lll}
u_i^0 = u_0, \quad & v_j^0\left(t\right) = v_0 \cos\left(\omega t + \varphi_j\right) \label{DielectricTop1} \\[6pt]
& w_j^0\left(t\right) \equiv v_{j+3}^0\left(t\right) = v_0 \sin\left(\omega t + \varphi_k\right), \qquad \label{DielectricTop2}
\end{IEEEeqnarray}
where $(u_0,v_0)$ are the critical points of the axially symmetric potential \eqref{AxiallySymmetricPotential} that satisfy \eqref{AxiallySymmetricExtrema3}, \eqref{AngularVelocity}. We set:
\begin{IEEEeqnarray}{ll}
u_i = u_i^0 + \delta u_i\left(t\right), \quad & v_i = v_i^0\left(t\right) + \delta v_i'\left(t\right) \\[6pt]
& w_i = w_i^0\left(t\right) + \delta w_i'\left(t\right). \qquad \label{AxialPerturbations}
\end{IEEEeqnarray}
\indent By plugging \eqref{DielectricTop1}--\eqref{AxialPerturbations} into the equations of motion \eqref{xEquation}--\eqref{yEquation} and using the minimization condition \eqref{AxiallySymmetricExtrema3} (for $u_0 \neq 0$), we obtain a second order system of linearized equations with periodic coefficients. Following \cite{AxenidesFloratosPerivolaropoulos00, AxenidesFloratosPerivolaropoulos01}, we may transform it into a second order system of constant coefficients by making an appropriate rotation in $SO(6)$. We get:
\begin{IEEEeqnarray}{ll}
\left[\begin{array}{c} \delta\ddot{\textbf{u}} \\ \delta\ddot{\textbf{v}} \\ \delta\ddot{\textbf{w}} \end{array}\right] + 2 \, \omega &\left[\begin{array}{ccc} 0 & 0 & 0 \\ 0 & 0 & -I_3 \\ 0 & I_3 & 0 \end{array}\right] \cdot \left[\begin{array}{c} \delta\dot{\textbf{u}} \\ \delta\dot{\textbf{v}} \\ \delta\dot{\textbf{w}} \end{array}\right] + \nonumber \\[6pt]
& \hspace{.3cm} + \left[\begin{array}{ccc} A_1 & A_2 & 0 \\ A_2 & B_1 & 0 \\ 0 & 0 & 0 \end{array}\right] \cdot \left[\begin{array}{c} \delta\textbf{u} \\ \delta\textbf{v} \\ \delta\textbf{w} \end{array}\right] = 0, \qquad \quad \label{LinearizedSystem1}
\end{IEEEeqnarray}
where
\begin{IEEEeqnarray}{l}
A_1 = u_0 \, I_3 + u_0\left(2u_0-1\right) \cdot \mathfrak{g}, \\[6pt]
A_2 = 2u_0v_0 \cdot \mathfrak{g}, \qquad \& \qquad B_1 = 2v_0^2 \cdot \mathfrak{g}, \label{LinearizedSystem2} \qquad
\end{IEEEeqnarray}
$I_3$ is the $3$-dimensional identity matrix and
\begin{IEEEeqnarray}{lll}
\mathfrak{g} \equiv
\left(\begin{array}{ccc}
0 & 1 & 1 \\
1 & 0 & 1 \\
1 & 1 & 0
\end{array}\right).
\end{IEEEeqnarray}
\indent In order to solve \eqref{LinearizedSystem1}--\eqref{LinearizedSystem2}, we plug the following general solution into \eqref{LinearizedSystem1}:
\begin{IEEEeqnarray}{l}
\left[\begin{array}{c} \delta\textbf{u} \\ \delta\textbf{v} \\ \delta\textbf{w} \end{array}\right] = \sum_{i=1}^{18} c_i \, e^{\lambda_i t} \, \boldsymbol{\xi}_i,
\end{IEEEeqnarray}
where the $c_i$ are constants determined by the initial conditions and $\lambda_i$, $\boldsymbol{\xi}_i$ solve the resulting eigenvalue problem for every $i = 1, \ldots 18$. A rather straightforward computation returns six zero eigenvalues (associated with the symmetries of the $SO\left(6\right)$ sector) and four nonzero eigenvalues: \\
\small\begin{IEEEeqnarray}{l}
\lambda_{1\pm}^2 = \frac{1}{9} - \frac{5u_0}{2} \pm \sqrt{\frac{1}{9^2} - \frac{u_0}{9} - \frac{5u_0^2}{12} + 4 u_0^3}, \label{AxiallySymmetricEigenvalues1} \\[6pt]
\lambda_{2\pm}^2 = \frac{5}{18} - \frac{5u_0}{2} \pm \sqrt{\frac{5^2}{18^2} - \frac{35u_0}{18} + \frac{163u_0^2}{12} - 20u_0^3}, \qquad \quad \label{AxiallySymmetricEigenvalues2}
\end{IEEEeqnarray}\normalsize \\
of multiplicities four and two respectively (so that $6 + 2\cdot4 + 2\cdot2 = 18$). A plot of the squares of the eigenvalues \eqref{AxiallySymmetricEigenvalues1}--\eqref{AxiallySymmetricEigenvalues2} as a function of the $SO\left(3\right)$ coordinate $u_0$ appears in figure \ref{Graph:Eigenvalues}. \\[6pt]
\indent In the allowed region \eqref{ExtremalBounds}, the spectrum of the axially symmetric configuration \eqref{DielectricTop1}--\eqref{DielectricTop2} always possesses 3 purely imaginary eigenvalues (for which $\lambda^2<0$) corresponding to stable directions. On the other hand, the square of the non-degenerate eigenvalue $\lambda_{2+}$ can either be positive or negative depending on whether $u_0$ is smaller or greater than $u_{\text{crit}}$. For $u_0 = u_{\text{crit}}$, $\lambda_{2+}^2$ flips sign making the corresponding direction change from stable ($\lambda_{2+}^2<0$) to unstable ($\lambda_{2+}^2>0$). Therefore the rightmost critical points ($u_0 > u_{\text{crit}}$) are always stable, whereas the leftmost ones ($u_0 < u_{\text{crit}}$) are unstable.\\[6pt]
\indent As it turns out, the same conclusion about the stability of \eqref{DielectricTop1}--\eqref{DielectricTop2} could have been reached had we perturbed the equations of motion \eqref{AxiallySymmetricExtrema1}--\eqref{AxiallySymmetricExtrema2}. The difference in this case is that there are only two fluctuation modes instead of nine and the angular momentum is essentially kept constant during the perturbation. The corresponding eigenvalues are given by $\lambda_{2\pm}$ in \eqref{AxiallySymmetricEigenvalues2}, giving rise to the same spectrum that we described in the previous paragraph. This result is of course consistent with the analysis of the eigenvalues of the Hessian matrix \eqref{HessianMatrix} that was presented at the end of section \ref{Section:ParticularSolutions}.
\section[Acknowledgements]{Acknowledgements}
The authors would like to thank Yuhma Asano, David Berenstein, Christos Efthymiopoulos, Jens Eggers, Jens Hoppe, Bum-Hoon Lee, Stam Nicolis and Georgios Pastras for illuminating discussions. E.F.\ and G.L.\ kindly acknowledge Luis \'{A}lvarez-Gaum\'{e} and the CERN Theory Group for instructive discussions and generous support. G.L.\ is grateful to Jens Hoppe and the KTH Royal Institute of Technology for hospitality and support during the early stages of this work. G.L.\ is also grateful to Charlotte Kristjansen and the Niels Bohr Institute and to Konstantinos Zoubos and the University of Pretoria for their hospitality and support.
\appendix
\bibliography{HEP_Bibliography,Math_Bibliography}
\end{document}